# Experimental and model-assisted analysis of lamella thinning and breakup in diesel-surrogate fuel-droplet wall impingement under Leidenfrost conditions

Xinquan Liang[a], Daniel Olufemi Olurotimi[a], Xi Liu[a], Minghui Xu[a], Yuhao Xu[a,*]

[a]Department of Mechanical Engineering, Clemson University, Clemson, SC 29634, USA

*Corresponding author. E-mail: yuhaox@clemson.edu

# Abstract

Hot-wall fuel-droplet impingement affects liquid redistribution, secondary droplet formation, and droplet evaporation in diesel-relevant spray-wall systems. This study investigates the spreading and breakup of single *n*-hexadecane droplets, used as a single-component diesel surrogate, on a heated stainless-steel wall at 300–500 °C over a Weber number ($We$) range of 5.77–208.61. High-speed backlit images were recorded during experiments and used to classify deposition, rebound, ejection, fragmentation, and splashing regimes, and to measure spreading-factor histories. The measured spreading histories were compared with a lamella-rim model to evaluate its predictive capability before droplet breakup and to infer the lamella state at experimentally observed breakup instants. The 300 °C cases did not enter the Leidenfrost regime under the present impact conditions and therefore serve as a non-Leidenfrost reference, deviating from the model predictions. For Leidenfrost cases at 350–500 °C, the model captures the pre-breakup spreading trajectory, including higher-$We$ cases that later undergo breakup. Wall temperature has a limited influence on the early spreading stage but more strongly affects later breakup timing after lamella thinning, especially in the intermediate-$We$ regime. At high $We$, breakup becomes increasingly inertia-dominated. Model-inferred lamella thicknesses evaluated at experimentally observed breakup instants are mainly concentrated between 0.010 and 0.017 of the initial droplet diameter. These results suggest that model-inferred lamella thickness can complement conventional Weber-number and temperature-based regime maps by providing local-state information for breakup timing in hot-wall fuel-droplet impingement models.

**Nomenclature**

| | |
|---|---|
| $a_R(t)$ | Rim cross-sectional radius, m |
| $D_0$ | Initial droplet diameter, m |
| $D_{\text{app}}(t)$ | Apparent near-wall contact diameter, m |
| $D_{\text{sp}}(t)$ | Overall spreading diameter, m |
| $F_{ilR}$ | Inertial force associated with the liquid entering the rim |
| $F_v$ | Viscous force at the lamella–rim interface |
| $F_\sigma$ | Capillary force due to surface tension |
| $h_{\text{lam}}(r,t)$ | Local liquid lamella thickness, m |
| $\overline{h^*_{R,Bk}}$ | The average $h^*_R$ value at the breakup instants |
| $h_R(t)$ | Lamella thickness at the lamella–rim junction, m |
| $R_R(t)$ | Rim-center radius, m |
| $R_{\text{sp}}(t)$ | Overall spreading radius, m |
| $\dot{R}_R(t)$ | Rim-center radial velocity, m/s |
| $\ddot{R}_R(t)$ | Rim-center radial acceleration, m/s$^2$ |
| $T_w$ | The temperature of the heated wall (which is also referred to as a heated surface), °C |
| $T_{sat}$ | Saturation temperature, °C |
| $r$ | Radial coordinate, m |
| $t$ | Physical time, s |
| $t_0$ | Time at the onset of the lamella–rim stage |
| $t_{end}$ | End time of the lamella stage |
| $t_{R,max}$ | Time at maximum spreading |
| $U_0$ | Impact velocity, m/s |
| $u_{\text{rim}}$ | Radial velocity of liquid entering the rim, m/s |
| $W_R(t)$ | Rim volume, m$^3$ |
| $We$ | Weber number |
| $Re$ | Reynolds number |

*Greek Letters*

| | |
|---|---|
| $\alpha$ | Velocity-enhancement factor |
| $\beta$ | Spreading factor |
| $\eta$ | Lamella thickness profile parameter |
| $\tau$ | Dimensionless time-shift parameter |
| $\rho_l$ | Liquid density, kg/m$^3$ |
| $\mu_l$ | Dynamic viscosity, Pa·s |
| $\nu_l$ | Kinematic viscosity, m$^2$/s |
| $\sigma_l$ | Surface tension, N/m |
| $\theta$ | Effective droplet-wall contact angle used in Eq. (24), ° |

*Superscripts*

| | |
|---|---|
| * | Nondimensionalized parameters |

*Subscripts*

| | |
|---|---|
| *app* | Apparent |
| *Bk* | Breakup instant |
| *end* | End state |
| *l* | Liquid |
| *lam* | Lamella |
| max | Maximum |
| ref | Reference |
| *sat* | Saturation |
| *sp* | Spreading |
| *w* | Wall |
| 0 | Initial state |

# 1. Introduction

Fuel-droplet wall impingement is a recurring process in direct-injection diesel engines, especially when spray droplets impinge on high-temperature surfaces such as cylinder walls and piston crowns during atomization, spray penetration, and near-wall transport [1-4]. Under sufficiently high wall-temperature conditions, these interactions can enter the Leidenfrost regime, which changes droplet spreading, rebound, and breakup behavior [5-9]. Because these interactions affect spray-wall dynamics, near-wall liquid-film formation, secondary droplet transport, and in-cylinder fuel redistribution, they can strongly influence fuel evaporation, mixing, combustion, engine performance, and emissions [1-4]. Directly studying these phenomena using practical diesel fuels is challenging because diesel is a complex multicomponent liquid whose thermophysical properties can change during heating and impact, making it difficult to isolate the local dynamics that control wall-impingement breakup [10-14]. The engineering problem addressed here is therefore to characterize breakup timing in a simplified diesel-surrogate hot-wall impingement system, with the goal of providing useful local-state information for spray-wall interaction modeling.

In this context, the present study uses a single-component surrogate fuel to avoid the added complexity of multicomponent fuels and to better control the experimental conditions. Specifically, *n*-hexadecane is used in this work because it is a common single-component surrogate fuel in diesel research [10-13, 15]. This choice enables simplified analysis of droplet impact under Leidenfrost conditions while retaining engineering relevance [15, 16]. The central question is whether existing approaches for characterizing droplet breakup on heated walls are sufficient to explain the timing and development of breakup during spreading.

Most existing descriptions of droplet breakup on heated walls rely on macroscopic parameters and global impact outcomes [8, 17-21], such as Weber number, wall temperature, and impact velocity. These studies often summarize droplet behavior using regime maps that identify whether rebound, ejection, fragmentation, or splashing occurs [8, 17, 21]. Such approaches are useful for determining whether breakup occurs, but they provide limited information on when breakup begins during spreading and which local dynamic phenomena govern it [8, 21]. This gap is consequential for engineering models of spray-wall interaction because secondary droplet generation depends not only on whether breakup occurs, but also on when it occurs relative to spreading, recoiling, evaporation, and near-wall fuel redistribution. A local-state indicator for breakup onset is therefore needed to connect single-droplet impact physics with practical wall-impingement modeling.

Reported experiments support the need for a local-state perspective. Although the impact force reaches its maximum when the droplet first contacts the wall, breakup does not necessarily occur at that moment and often develops later during the spreading stage [18, 21-23]. This suggests that breakup onset is governed not only by the initial impact force but also by the subsequent evolution of the liquid film during spreading. Under Leidenfrost conditions, the presence of a vapor layer significantly alters the near-wall flow and liquid-film dynamics [6, 24, 25]. As a result, local parameters such as central liquid-film thickness (i.e., lamella thickness), spreading velocity, and rim development may affect the breakup process [22, 23, 26, 27]. However, compared with macroscopic outcome-based descriptions and regime classification, the connection between lamella-scale evolution and breakup timing remains insufficiently resolved, especially for diesel-surrogate hot-wall impingement systems [22, 23, 25, 27]. This limits the current understanding of Leidenfrost droplet impact and its use in engineering models of hot-wall fuel-droplet impingement.

To connect global impact outcomes with breakup development during spreading, a theoretical framework is needed to describe the droplet's continuous pre-breakup motion. The lamella-rim model proposed by Castanet et al. [26] provides such a reference for droplet spreading under Leidenfrost conditions. In this model, the spreading droplet is represented as a continuous, axisymmetric system comprising a central thin lamella and a peripheral rim. Conservation of mass and momentum in the rim region is then used to predict the time-resolved evolution of the spreading diameter. Unlike regime maps or maximum-spreading correlations, this model provides a continuous spreading trajectory and a corresponding estimate of lamella evolution, making it useful for analyzing local liquid-film states during hot-wall droplet impingement.

The model reported in Ref. [26] was not developed as a breakup model and assumes that the droplet remains continuous during spreading. Therefore, it cannot directly predict breakup onset or post-breakup fragmentation. In the present work, the model [26] is instead used as a model-assisted reconstruction of the continuous pre-breakup spreading trajectory and the corresponding lamella state. By mapping experimentally observed breakup instants onto the model-inferred lamella-thickness evolution, this study evaluates whether breakup timing can be interpreted more clearly using the local thinning state of the lamella than using global impact parameters alone. In this way, the model is used not as a complete breakup predictor, but as a diagnostic framework for identifying breakup-relevant local states in diesel-surrogate hot-wall fuel-droplet impingement.

Accordingly, this paper investigates single *n*-hexadecane droplets impacting a heated stainless-steel wall as a simplified diesel-surrogate wall-impingement system. High-speed imaging is used to classify impact outcomes and measure spreading histories over wall temperatures of 300–500 °C and Weber numbers of approximately 5.77–208.61. The measured spreading histories are compared with the lamella-rim model [26] to evaluate continuous pre-breakup spreading and to estimate the lamella state at experimentally observed breakup instants. The objective is to determine whether model-inferred lamella thinning can serve as a local-state indicator of breakup timing under Leidenfrost wall-impingement conditions. The results provide a model-assisted interpretation of secondary breakup in hot-wall fuel-droplet impingement and offer local-state guidance for future engineering spray-wall interaction models.

# 2. Experimental methods

## 2.1 Experimental setup

Experiments reported in this work are conducted to investigate the spreading and breakup of single *n*-hexadecane (boiling point: 287 °C at 1 atm [28]) droplets impacting a heated stainless-steel wall. The setup is designed to ensure stable droplet generation, controllable impact conditions, a well-defined wall temperature, and high-speed imaging of the impact process.

As shown in Fig. 1, *n*-hexadecane is delivered to the needle by a syringe pump, which slowly feeds the liquid to allow a droplet to grow at the needle tip until it detaches naturally under gravity. This quasi-static feeding process produces droplets with repeatable diameters. Repeated tests show that the droplet diameter remains within the range of 1.897–2.204 mm. Most of the initial droplet diameters fall within the range of 1.90 to 2.05 mm. The release height is set using a height gauge, and different release heights are used to obtain different impact velocities. After detachment, the droplet falls vertically and impacts the stainless-steel plate affixed to the electric hot plate. In this setup, the hot plate serves as the heat source, while the stainless-steel plate is the actual impingement surface.

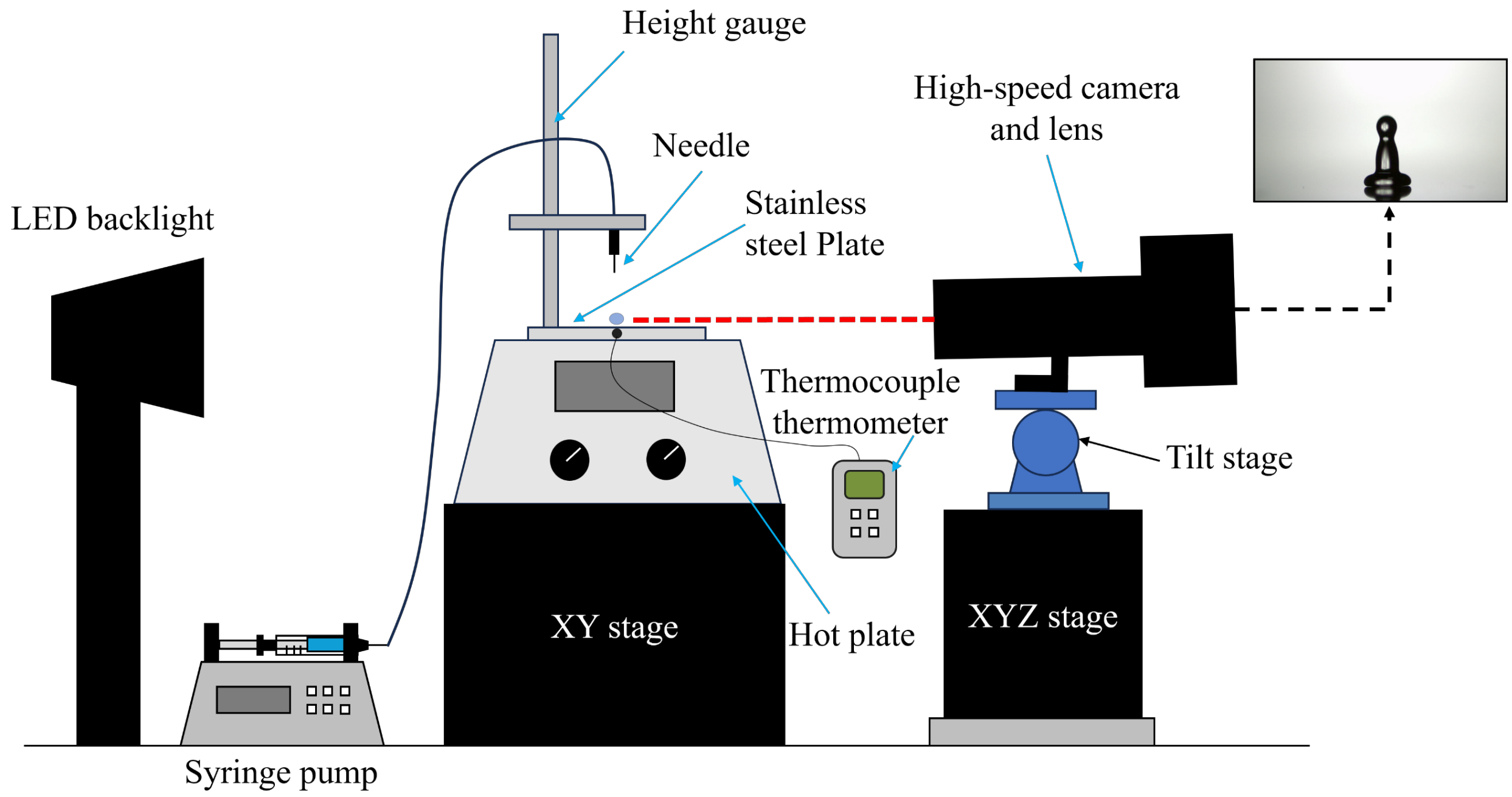

**Fig. 1.** Schematic of the experimental setup used in this study.

Type 304 stainless steel was chosen as the plate material because of its superior thermal and structural stability at high temperatures. The hot plate used in this study can heat the surface of the 304 stainless steel plate to over 500 °C. Under such conditions, copper plates are found to oxidize rapidly and darken at around 400 °C, while the surfaces of aluminum plates deform noticeably when the temperature exceeds 500 °C. By comparison, the stainless-steel plate can maintain an undeformed mirror-like surface even at temperatures above 500 °C, making it more suitable for repeatable high-temperature droplet-impact experiments. In the experiments, a 304 stainless-steel plate measuring 152.4 mm × 152.4 mm × 3.18 mm (6 in × 6 in × 1/8 in) is used as the impingement surface. The upper surface of the plate is ground and polished to a mirror finish and serves as the heated wall for inducing the Leidenfrost effect.

The wall temperature of the stainless-steel plate is monitored by a thermocouple attached to the plate. Before each test, the plate is heated to the prescribed temperature and held until a stable thermal condition is reached. The droplet is then released from the needle to impact the heated wall.

The impact process is recorded by a Kron Chronos 4K12 high-speed camera equipped with a Nikon 200 mm Macro lens and positioned horizontally with respect to the test section. The camera captures 1280 × 720-pixel videos at 5153 frames per second (fps). An LED backlight is placed opposite the camera to provide high-contrast shadow images of the droplet profile. The camera is mounted on an XYZ stage and a tilt stage to facilitate precise alignment and focusing. The recorded backlit images are used for subsequent measurements of the initial droplet diameter, impact velocity, spreading diameter, and breakup time.

### 2.2 Experimental matrix and droplet behavior definitions

Using the experimental setup described in Section 2.1, droplet impact experiments are carried out under different wall temperatures and impact conditions. The results show that the droplet can exhibit several distinct post-impact behaviors under different initial impact velocities and heated-wall temperatures. For clarity in the following discussion, the impact outcomes are classified as deposition, rebound, ejection, fragmentation, and splashing. As shown in Fig. 2, the 300 °C cases correspond to deposition and exhibit non-Leidenfrost behavior [29]; therefore, they are classified as non-Leidenfrost cases based on the observed impact outcome, whereas the 350–500 °C cases exhibit Leidenfrost-related outcomes including rebound, ejection, fragmentation, and splashing.

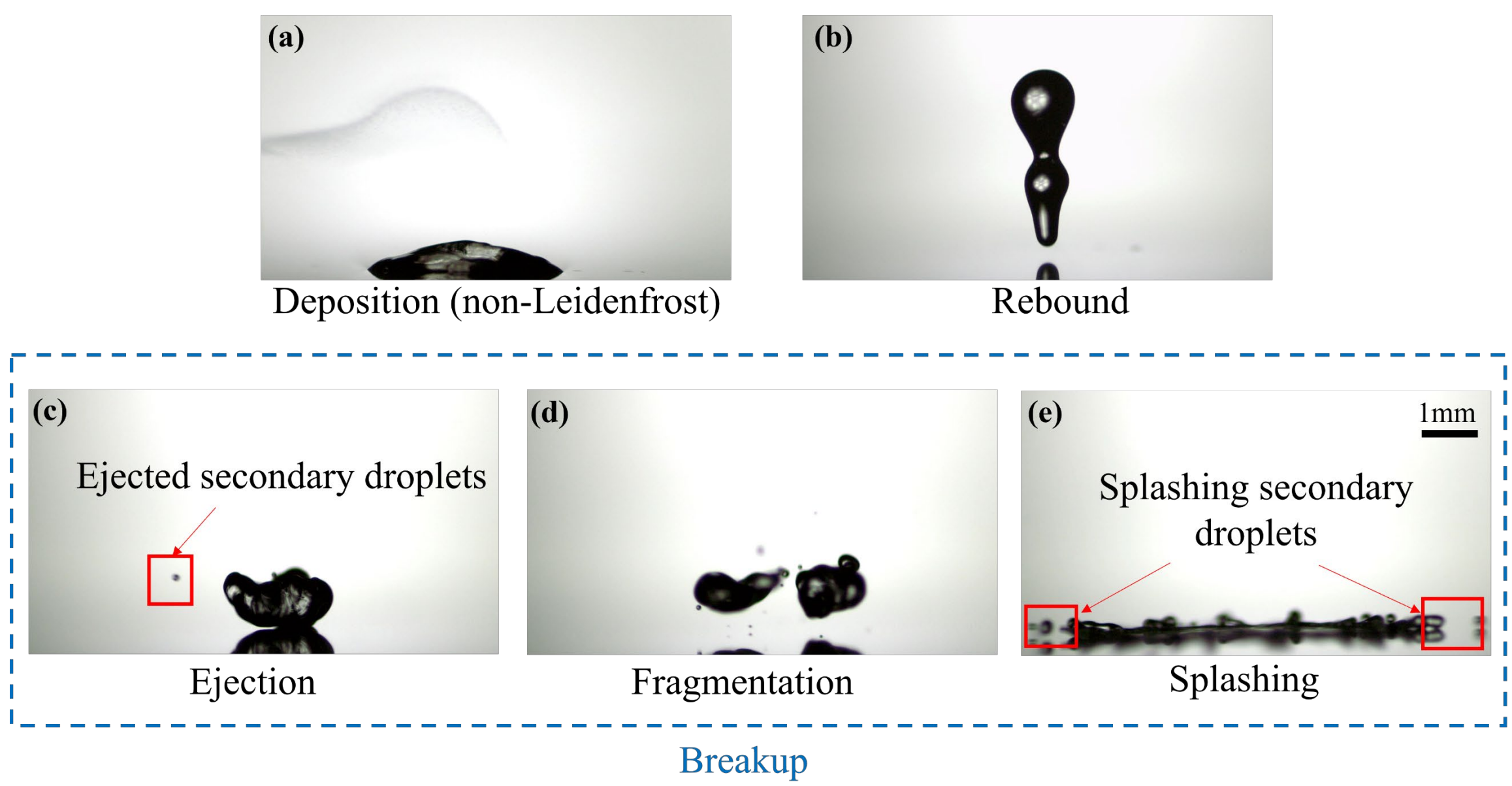

**Fig. 2.** Different droplet behaviors after impact on the heated wall under different wall temperatures. (a) Deposition [29]: under non-Leidenfrost conditions, the droplet spreads and subsequently recedes while remaining in contact with the heated wall, without rebound; (b) Rebound [23]: The droplet bounces off the heated wall after the impact; (c) Ejection [17]: a small secondary droplet is ejected from the impacting droplet; (d) Fragmentation [19]: the impacting droplet breaks into several smaller droplets; (e) Splashing [30]: the impacting droplet splashes into many very small secondary droplets. The blue dashed box indicates three behaviors that are categorized as the "breakup" behavior in the subsequent discussion.

In the deposition regime, the droplet spreads and subsequently recedes while remaining in contact with the heated wall, without complete rebound. This definition follows the outcome-based terminology in which deposition includes drop spreading and receding and is distinguished from rebound by the absence of post-impact bouncing [29]. In the rebound regime, the droplet does not break up and bounces off the heated wall without fracturing or ejecting smaller droplets. The classification of this rebound regime is based on previously published work [23]. In the ejection regime, one or several small secondary droplets are emitted during the retraction stage, while the rest of the main liquid body remains intact for most of the impact process [17]. The retraction stage here refers to the period after the droplet reaches its maximum diameter, during which the rim moves inward under surface tension. For Leidenfrost cases, this retraction occurs on a vapor layer and may precede rebound or breakup [30]; for the 300 °C non-Leidenfrost cases, the droplet recedes while remaining deposited on the wall [29]. In the fragmentation regime, the retracting liquid structure breaks into several separated liquid masses [19]. In the splashing regime, the droplet shatters into many small secondary droplets during the early spreading stage, making this the most dramatic impact response among the five cases. This splashing regime is consistent with that defined in prior studies [30].

It should be noted that these four Leidenfrost-related outcome labels are not always strictly exclusive in the experiments.  More than one behavior can occur in the same impact case. In this study, each Leidenfrost case is classified according to the first clearly identifiable post-impact behavior captured in the high-speed image sequence. For example, in a fragmentation case, the droplet first breaks into several separated liquid masses. Some of these secondary droplets may then eject smaller droplets or rebound from the heated wall. This phenomenon can be observed in Fig. 3.

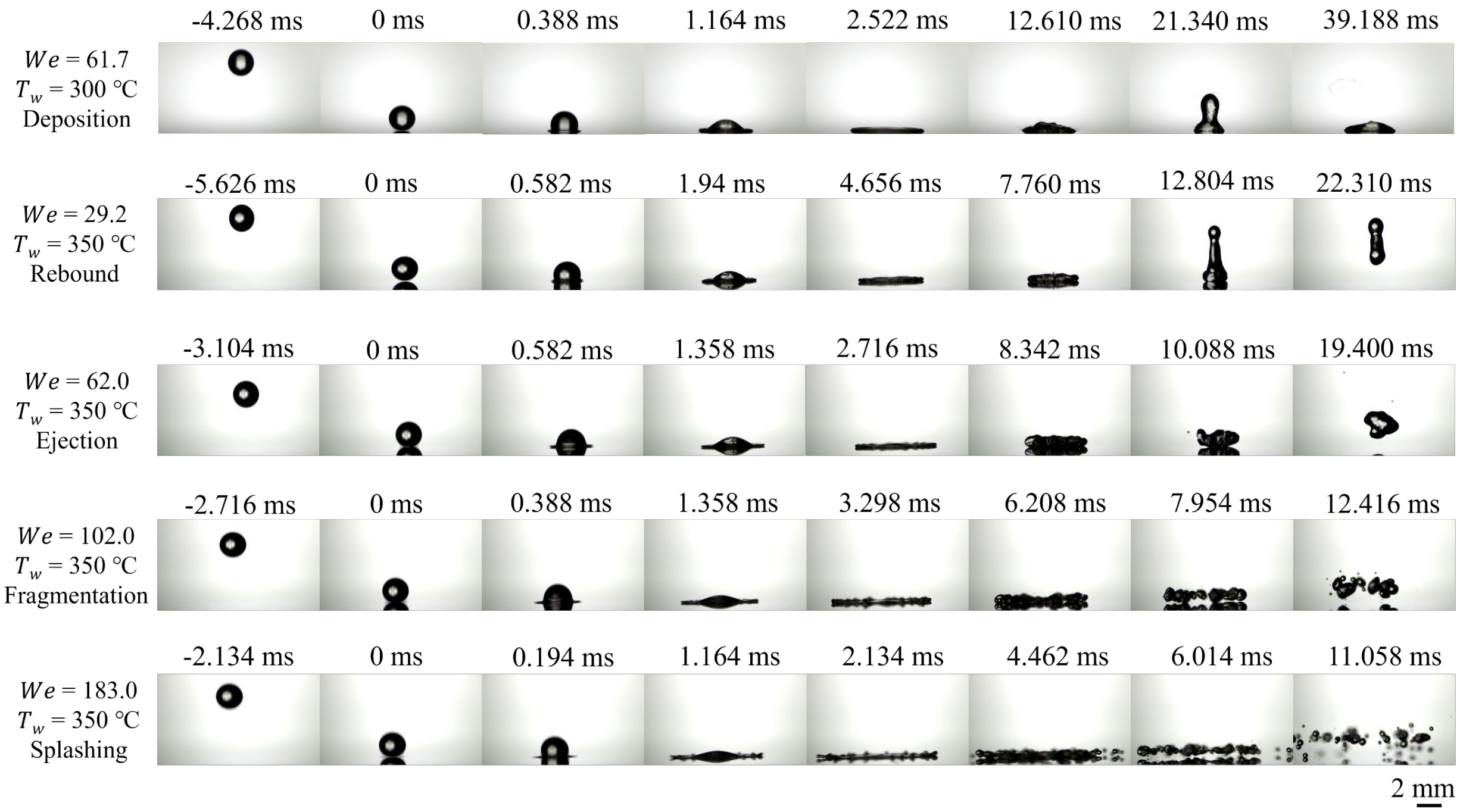

**Fig. 3.** Sequential images of representative *n*-hexadecane droplet impacts. The upper row shows a deposition case at $We$ = 61.7 and $T_w$ = 300 °C under non-Leidenfrost conditions. The lower four rows show Leidenfrost cases at $T_w$ = 350 °C, where the outcome changes from rebound to ejection, fragmentation, and splashing as $We$ increases from 29.2 to 183.0.

Fig. 3 shows representative time-resolved image sequences of n-hexadecane droplets impacting the heated wall at $T_w = 300$ and 350 °C. The figure summarizes the different droplet behaviors observed after impact under different wall temperature conditions. It also shows how the impact outcome changes as the Weber number increases. For the non-Leidenfrost case ($T_w =$ 300 °C, $We = 61.7$), the droplet spreads and recedes while remaining on the wall without rebound. For the Leidenfrost cases, at a low Weber number ($We = 29.2$), the droplet first spreads on the heated wall. It then retracts and rebounds from the heated wall. No breakup or secondary droplet formation is observed in this case. At a moderately low Weber number ($We = 62.0$), the droplet also first spreads and then retracts. However, during the rebound process, one or several small secondary droplets are ejected from the main droplet. The main liquid body remains mostly intact. At an intermediate Weber number ($We = 102.0$), the droplet first spreads over the heated wall. It then breaks into several relatively large secondary droplets during the later stage. These secondary droplets also rebound and leave the heated wall. At a high Weber number ($We = 183.0$), the droplet behavior becomes more dramatic. Many small secondary droplets are splashed out during the early spreading stage. After this splashing process, the remaining main liquid body further fragments into many larger secondary droplets. Since the entire process of splashing may involve two distinct breakup behaviors—ejection and fragmentation—this study defines ejection and fragmentation as transitional behaviors in the evolution of droplet breakup toward splashing. The lower four image sequences in Fig. 3 show the evolution of Leidenfrost impact behavior from rebound to ejection, fragmentation, and splashing with increasing impact velocity.

Fig. 4 summarizes the experimental matrix and the corresponding impact behaviors defined in Fig. 2. Because the dynamic Leidenfrost temperature depends on impact conditions, surface properties, and fluid properties [25], no fixed Leidenfrost-point value is used here; the Leidenfrost or non-Leidenfrost cases are classified based on the observed impact outcomes. Experiments are conducted at five wall temperatures, namely 300, 350, 400, 450, and 500 °C. For each wall-temperature condition, the droplet release height is adjusted to vary the impact velocity $U_0$, thereby producing a series of impact conditions. These impact conditions are characterized using the Weber number and Reynolds number, which represent the relative importance of inertial effects relative to surface-tension and viscous effects, respectively:

$$We = \frac{\rho_l D_0 U_0^2}{\sigma_l} \tag{1}$$

$$Re = \frac{\rho_l D_0 U_0}{\mu_l} \quad (2)$$

In total, 100 cases are tested, with 20 cases at each wall-temperature condition. The symbol assigned to each case denotes the corresponding impact outcome according to the regime classification defined in Fig. 2.

As shown in Fig. 4, the impact outcome depends on both wall temperature and Weber number. At 300 °C, all tested cases correspond to deposition: the droplets spread and recede while remaining on the heated wall without rebound. For the 350–500 °C cases, the impact outcome changes progressively from rebound to ejection, fragmentation, and splashing as the Weber number increases. These results indicate the importance of impact inertia in controlling the impact behaviors, which motivates the detailed analysis discussed in the subsequent sections.

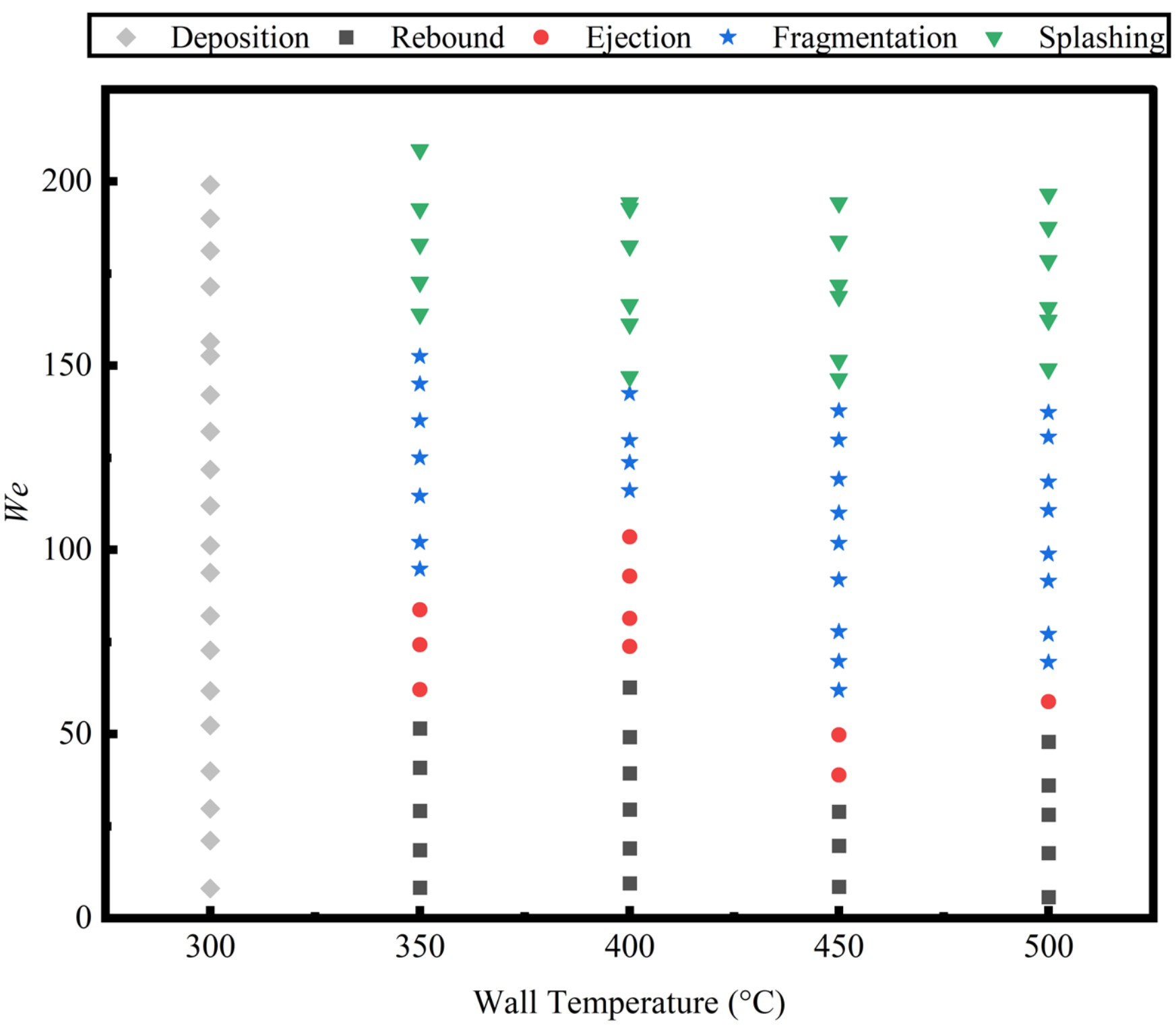

**Fig. 4.** Regime map of all experiments conducted in this work, highlighting five different regimes.

### 2.3 Parameter definitions

Fig. 5 illustrates the morphological evolution of an *n*-hexadecane droplet impacting a superheated wall in the Leidenfrost regime, together with the geometric parameters used in the subsequent analysis. The initial droplet diameter is denoted by $D_0$. Variables marked with an asterisk ($*$) represent dimensionless quantities. In the present study, length-related quantities are normalized by the initial droplet diameter $D_0$, and time is normalized by the characteristic inertial timescale $D_0/U_0$, where $U_0$ is the impact velocity. Accordingly, $t^*$, $t_0^*$, $t_{R,\mathrm{max}}^*$, $t_{\mathrm{end}}^*$, $R_{\mathrm{sp}}^*$, $R_R^*$, and $a_R^*$ denote the corresponding dimensionless time and length parameters, which are defined below:

$$t^* = \frac{tU_0}{D_0} \tag{3}$$

$$t_0^* = 0 \tag{4}$$

$$t_{R,max}^* = \frac{t_{R,max}U_0}{D_0} \tag{5}$$

$$t_{end}^* = \frac{t_{end}U_0}{D_0} \tag{6}$$

$$R_{sp}^*(t) = \frac{R_{sp}(t)}{D_0} \tag{7}$$

$$a_R^*(t) = \frac{a_R(t)}{D_0} \tag{8}$$

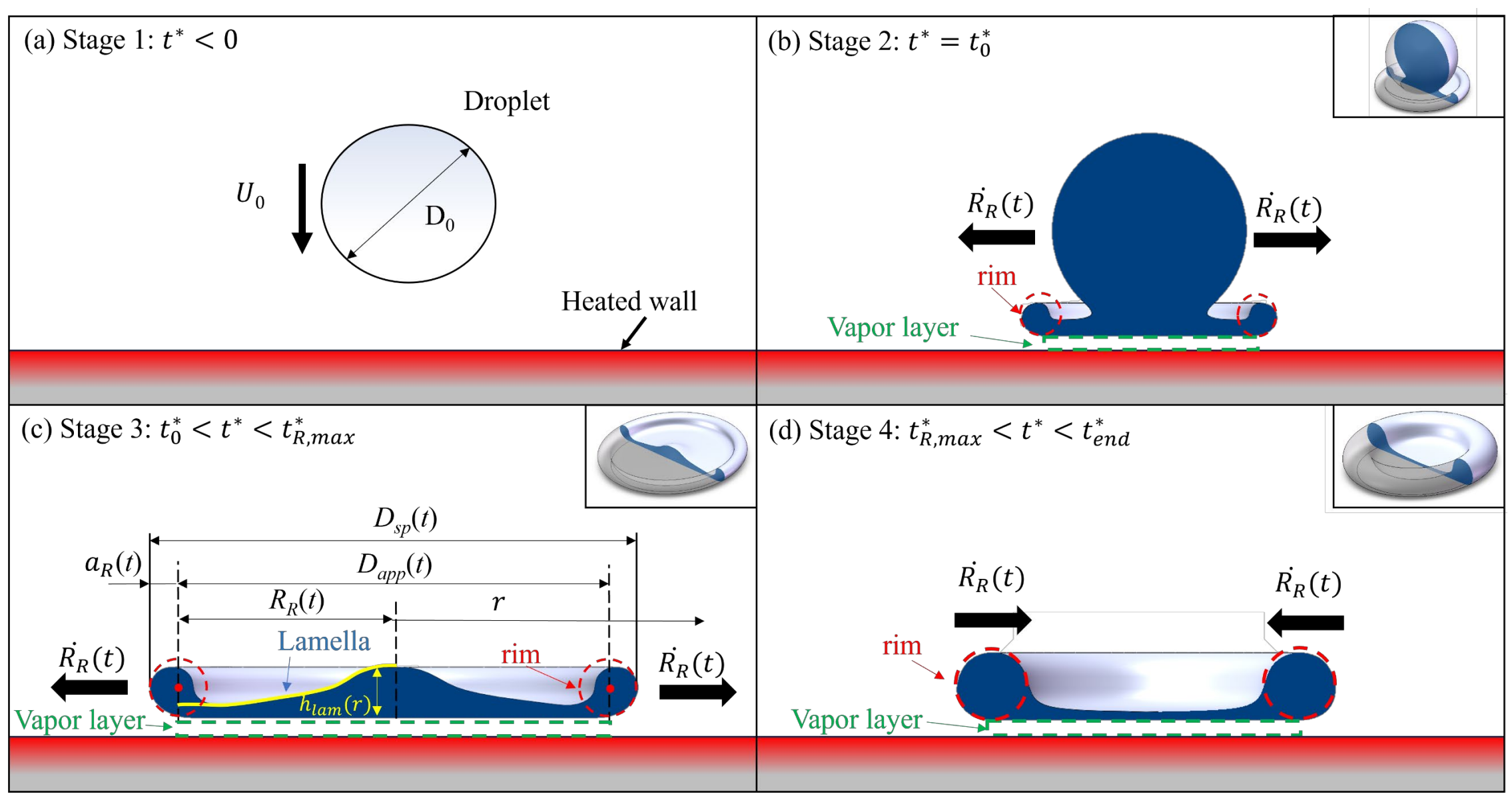

**Fig. 5.** Droplet impact deformation process in four stages and parameter definitions. (a) Stage 1: the droplet has not impacted the heated wall; (b) Stage 2: the droplet starts to deform and form the rim and lamella; (c) Stage 3: the droplet continues to deform and reaches its maximum spreading diameter; (d) Stage 4: the droplet starts to retract, i.e., the retraction stage discussed earlier.

Under Leidenfrost conditions, a vapor layer forms between the droplet and the heated wall. In Fig. 5(b), the droplet is in the early kinematic stage, where it starts to deform under impact inertia, and the rim first becomes identifiable. In Fig. 5(c), the droplet enters the spreading stage, during which a thin lamella expands radially, and a thicker rim develops at its edge. The time $t_{R,\mathrm{max}}^*$ corresponds to the moment of maximum spreading. In Fig. 5(d), the droplet enters the retraction stage, where the rim retracts toward the center under surface tension. During this stage, however, the rim volume continues to increase because liquid is still transferred from the lamella into the rim, and the lamella thickness continues to decrease rather than being replenished. The time $t_{\mathrm{end}}^*$ indicates the time when the lamella disappears.

As defined in Fig. 5(c), $R_{\mathrm{sp}}$ is the distance from the droplet centerline to the outer edge of the rim, $R_R$ is the distance from the centerline to the rim center, and $a_R(t)$ is the rim radius. The lamella thickness is denoted by $h_{lam}(r,t)$, which varies with the radial position $r$. The spreading velocity is defined as $\dot{R_R} = dR_R/dt$. These parameters are used to characterize the droplet evolution from the onset of rim formation, through radial spreading, to the final retraction process.

### 2.4 Measurement and uncertainty

The droplet impact process is recorded using a high-speed camera at 5153 fps with backlit imaging. The recorded image sequences are used to determine the initial droplet diameter $D_0$, impact velocity $U_0$, spreading diameter $D_{sp}(t)$, and breakup time $t_{Bk}$. Specifically, $D_0$ is measured from pre-impact droplet images, $U_0$ is obtained from the droplet displacement over known frame intervals before impact, $D_{sp}(t)$ is extracted from the time-resolved spreading profile, and the breakup time is manually identified from the first frame showing visible breakup.

Measurement uncertainties are evaluated for the image-based quantities used in this work, as summarized in Table 1. The image-to-physical length calibration relative standard uncertainty is estimated from repeated manual measurements of the calibration tool as 0.81%. This calibration uncertainty is combined with the image-segmentation uncertainty to estimate the standard uncertainties of $D_0$ and $D_{sp}(t)$, giving $u_{D_0,total} \approx 0.023$ mm and $u_{D_{sp}(t),total} = 0.3423$ mm. The uncertainty in $\beta(t) = D_{sp}(t)/D_0$ is obtained by standard uncertainty propagation, giving $u_\beta = 0.1750$, corresponding to a relative standard uncertainty of 3.33%. The relative standard uncertainty of $U_0$ is 0.81–0.84%, which propagates to a 1.62–1.68% relative uncertainty contribution in $We$ because $We \propto U_0^2$. Including both $D_0$ and $U_0$, and neglecting the uncertainties in $\rho_l$ and $\sigma_l$, the combined relative standard uncertainty of $We$ is approximately 2.0%. For the Reynolds number, the uncertainty contribution from $U_0$ alone is therefore 0.81–0.84%. When both $D_0$ and $U_0$ are included, and the uncertainties in $\rho_l$ and $\mu_l$ are neglected, the combined relative standard uncertainty of $Re$ is approximately 1.4%, since $Re \propto \rho_l D_0 U_0/\mu_l$. The heated-wall temperature $T_w$ is measured using the temperature-control instrumentation, and its standard instrumental uncertainty is estimated as $u_{T_w,\mathrm{instr}} = 0.74\text{–}1.20\ °\mathrm{C}$, depending on the wall-temperature condition. The breakup time is manually determined from the high-speed image sequences, and the uncertainty associated with frame selection is estimated as 0.112 ms based on a $\pm 1$ frame selection uncertainty for both the impact and breakup frames. For the model-inferred lamella thickness at breakup, $h^*_{R,Bk}$, the relative combined standard uncertainty is evaluated by propagating the relevant input uncertainties through the lamella-rim model at the experimentally identified breakup instants. The median relative uncertainty is approximately 1.8%. The uncertainty propagation procedure used here follows standard first-order propagation of the measured input quantities and the model-inferred lamella thickness.

**Table 1**

Summary of measurement uncertainty estimates used in the present analysis.

| **Quantity** | **Symbol** | **Standard uncertainty used in this work** |
|---|---|---|
| Image-to-physical length calibration factor | $\varepsilon$ | $u_\varepsilon/\varepsilon$ = 0.81% relative standard uncertainty |
| Initial droplet diameter | $D_0$ | $u_{D_0,total} \approx 0.023$ mm |
| Spreading diameter | $D_{sp}(t)$ | $u_{D_{\mathrm{sp}},\mathrm{total}}(t) = 0.3423$ mm |
| Spreading factor | $\beta(t)$ | $u_\beta = 0.1750$, corresponding to $u_\beta/\beta = 3.33\%$ |
| Impact velocity | $U_0$ | $u_{U_0}/U_0 = 0.81 - 0.84\%$ |
| Weber number, uncertainty from $U_0$ only | $We$ | $(\frac{u_{We}}{We})_{U_0} = 1.62 - 1.68\%$ |
| Weber number, uncertainty from $D_0$ and $U_0$ | $We$ | $\frac{u_{We}}{We} \approx 2\%$ |
| Reynolds number, uncertainty from $U_0$ only | $Re$ | $(\frac{u_{Re}}{Re})_{U_0} = 0.81 - 0.84\%$ |

| Quantity | Symbol | Standard uncertainty used in this work |
|---|---|---|
| Reynolds number, uncertainty from $D_0$ and $U_0$ | $Re$ | $\frac{u_{Re}}{Re} \approx 1.4\%$ |
| Breakup time | $t_{Bk}$ | $u_{t_{Bk}} = 0.112$ ms |
| Temperature of the heated wall | $T_w$ | $u_{T_w,\mathrm{instr}} = 0.74$–$1.20$ °C |
| Model-inferred breakup lamella thickness | $h^*_{R,Bk}$ | $\left[\frac{u_c\left(h^*_{R,Bk}\right)}{h^*_{R,Bk}}\right]_{\mathrm{med}} \approx 1.8\%$ |

## 2.5 Prediction model

The theoretical analysis in this study follows the analysis reported in Ref. [26], whose model was originally used to predict cases with low-$We$ values without droplet breakup. In this work, we extend the applicability of this model to higher-$We$ cases. The model proposed by Castanet et al. [26] is introduced below and then compared with our experimental data in Section 3.

The analysis starts from $t_0^*$ and continues until $t_{end}^*$ (i.e., stages 2–4 in Fig. 5). During this period, the post-impact droplet is simplified as an axisymmetric structure composed of a central thin liquid film (lamella) and an outer edge liquid ridge (rim). Another relevant parameter for the analysis is the droplet spreading factor $\beta$:

$$\beta(t) = \frac{D_{sp}(t)}{D_0} = \frac{2R_{sp}(t)}{D_0} = 2R_{sp}^*(t) = 2(R_R^* + a_R^*) \tag{9}$$

In addition to being measured based on dimensions obtained from the high-speed images, $\beta(t)$ can also be predicted theoretically. In the present work, this prediction is based on the rim-dynamics-based drop-spreading model proposed by Castanet et al. [26], which solves $R_R^*$ and $a_R^*$ for Eq. (9). The comparison between the measured and predicted spreading curves will provide the basis for the discussion in Section 3.

The core of this prediction model for obtaining $\beta(t)$ is to determine the rim motion from a momentum balance applied to the rim control volume (see the red dashed regions in Fig. 5(b)–(d)).

This momentum balance [26] is written as

$$\rho_l W_R \ddot{R}_R = F_{ilR} - F_v - F_\sigma \quad \text{at } r = R_R \tag{10}$$

This equation is essentially Newton's second law applied to the rim: the rim mass–acceleration term is balanced by the inertia of the liquid entering the rim, the viscous contribution at the lamella–rim interface, and the capillary resistance. Following Castanet et al., with the introduction of the velocity-enhancement factor $\alpha$ to account for the vapor-induced increase in the velocity of the liquid entering the rim [26], the nondimensionalized form of each term of Eq. (10) is expressed as [26]:

$$\left(\rho_l W_R \ddot{R}_R\right)^* = \frac{W_R^*}{2\pi R_R^* h_R^*} \ddot{R}_R^* \tag{11}$$

$$(F_{ilR})^* = \left(u_{\mathrm{rim}}^* - \dot{R}_R^*\right)^2 \tag{12}$$

$$(F_v)^* = \frac{6(1+\alpha)}{Re(t^* + \tau)} \tag{13}$$

$$(F_\sigma)^* = \frac{2}{We h_R^*} \tag{14}$$

As such, Eq. (10) can be nondimensionalized as [26]:

$$\frac{W_R^*}{2\pi R_R^* h_R^*} \ddot{R}_R^* = \left(u_{\mathrm{rim}}^* - \dot{R}_R^*\right)^2 - \frac{6(1+\alpha)}{Re(t^* + \tau)} - \frac{2}{We h_R^*} \tag{15}$$

where $\tau$ is a dimensionless time shift parameter used to correct the initial time point of the asymptotic solution during the early stage of spreading [26]. To close Eq. (15), the quantities $u_{\mathrm{rim}}^*$, $h_{lam}^*$, and $W_R^*$ must be expressed as functions of $R_R^*$ and $t^*$. Following Castanet et al. [26], the lamella thickness is described by the self-similar Gaussian profile [31, 32]:

$$h_{lam}^*(r^*, t^*) = \frac{\eta}{(t^* + \tau)^2} \exp\left[-\frac{6\eta (r^*)^2}{(t^* + \tau)^2}\right] \tag{16}$$

where $\eta$ is a dimensionless geometric parameter that describes the shape of the liquid-film thickness distribution, and it determines the central film thickness and the radial decay characteristics of the lamella. Following Castanet et al. [26], the present study takes $\eta = 0.39$ and $\tau = 0.25$.

At the lamella–rim junction ($R^* = R_R^*$), the local lamella thickness can be obtained by:

$$h_R^*(t^*) = \frac{\eta}{(t^* + \tau)^2} \exp\left[-\frac{6\eta (R_R^*)^2}{(t^* + \tau)^2}\right] \tag{17}$$

This expression indicates that the thickness of the liquid film decreases rapidly with time and is distributed in a Gaussian attenuation along the radial direction.

Regarding $u_{\mathrm{rim}}^*$ in Eq. (15), the following expression is provided by Castanet et al. [26]:

$$u_{\mathrm{rim}}^* = \frac{(1 + \alpha) R_R^*}{t^* + \tau}, \text{at } R^* = R_R^* \tag{18}$$

Finally, the rim volume needed for Eq. (15) is obtained from volume conservation. The liquid volume remaining outside the lamella region is assigned to the rim, giving [26]:

$$W_R^* = \frac{\pi}{6} - \int_0^{R_R^*} 2\pi\, R^* h_{lam}^*(R^*, t^*)\, dR^* \tag{19}$$

Note that there is a typo in this equation in Castanet et al. [26]. We have reworked the integration to confirm that Eq. (19) presented here is the correct form, which should be a nondimensionalized equation.

Substituting Eq. (16) into Eq. (19) and solving the integration yields:

$$W_R^* = \frac{\pi}{6} \exp\left[-\frac{6\eta (R_R^*)^2}{(t^* + \tau)^2}\right] \tag{20}$$

Then, substituting Eqs. (17), (18), and (20) into Eq. (15) gives a single second-order ordinary differential equation for $R_R^*(t^*)$:

$$\ddot{R}_R^* = \frac{12\eta R_R^*}{(t^* + \tau)^2} \left\{ \left[ \frac{(1 + \alpha) R_R^*}{t^* + \tau} - \dot{R}_R^* \right]^2 - \frac{6(1 + \alpha)}{Re(t^* + \tau)} \right\} - \frac{24 R_R^*}{We} \exp\left[\frac{6\eta (R_R^*)^2}{(t^* + \tau)^2}\right] \tag{21}$$

Eq. (21) is solved in MATLAB using the ode45 solver. The velocity-enhancement factor $\alpha$ is determined for each case by matching the predicted maximum spreading factor to the free-slip maximum-spreading correlation described below.

After $R_R^*(t^*)$ is obtained, $W_R^*(t^*)$ is calculated from Eq. (20). To get $a_R^*$ needed for Eq. (9), the rim is approximated as a torus with a circular cross-section; therefore,

$$W_R^* = (2\pi R_R^*) \times \pi (a_R^*)^2 \tag{22}$$

and the rim radius can be calculated as

$$a_R^* = \left(\frac{W_R^*}{2\pi^2 R_R^*}\right)^{1/2} \tag{23}$$

Equations (17), (20), and (23) show that $h_R^*$, $W_R^*$, and $a_R^*$ can all be determined once $R_R^*(t^*)$ is known. Once $a_R^*$ is obtained from $W_R^*$ and $R_R^*$, the predicted spreading factor $\beta(t)$ can be computed from Eq. (9).

To solve Eq. (21), the parameter $\alpha$ must be known. In Castanet et al. [26], $\alpha$ is adjusted by comparing the model prediction with experimental spreading curves. In the present work, a different approach is used to determine $\alpha$.

Here, a series of $\alpha$ values are used, starting with $\alpha = 0$, to generate $\beta(t)$ curves, from which $\beta_{\max}$ values can be obtained.

The value of $\alpha$ is determined such that the resulting $\beta(t)$ curve yields the same reference maximum spreading factor $\beta_{\max,\mathrm{ref}}$, which is estimated using the energy-based free-slip model proposed by Wildeman et al. [33]:

$$\beta_{\max,\mathrm{ref}} = \left[\frac{4}{1 - \cos\,\theta}\left(\frac{We}{24} + 1\right)\right]^{1/2} \tag{24}$$

In this work, $\theta$ is taken as 105° for the free-slip limit. The value of α is obtained iteratively by varying α until the predicted maximum spreading factor matches the reference maximum spreading factor defined above.

## 3. Results and model-assisted interpretation

To investigate how droplet spreading dynamics on a heated wall vary with impact intensity, and to evaluate the applicability of the model reported in Ref. [26] under different conditions, representative experimental results under different $We$ conditions and the temporal evolution of the droplet spreading factor, $\beta(t)$, are presented and discussed here. It should be emphasized that the focus of this work is not merely on whether the model can provide a reasonable spreading curve, but more importantly on the time at which the experimental curve begins to deviate from the model prediction at different wall temperatures, as well as the pattern of this deviation as $We$ changes. It should also be noted that, among the five wall-temperature conditions considered in this study, in the 300 °C cases, the droplets remain in direct contact with the heated wall throughout the spreading process, exhibiting behavior clearly distinct from the Leidenfrost effect. On this basis, the 300 °C condition in this study serves not only as a relatively low wall-temperature case, but also as a non-Leidenfrost control group for comparison with the Leidenfrost cases at 350–500 °C. The deposition behavior of 300 °C cases over the tested velocity range provides a baseline for confirming that the subsequent 350–500 °C cases are analyzed within the Leidenfrost-regime framework. For this reason, the discrepancy between the experimental curve and the model in Ref. [26] at 300 °C should not be simply interpreted as a model prediction error, but rather as the result of the inadequacy of the model for non-Leidenfrost impact. For each subplot in Figs. 6–8, the $\beta_{\text{Prediction}}$ was calculated using the average Weber and Reynolds numbers of the four Leidenfrost cases at 350–500 °C under the same nominal impact condition. The averaged $We$ and $Re$ values used in the calculation are reported in each subplot. Accordingly, the results in the low-$We$ (8–39), intermediate-$We$ (50–122), and high-$We$ (131–199) regimes are discussed separately below.

### 3.1 Spreading behavior in the low-$We$ regime

Under relatively low $We$ conditions, the Leidenfrost cases at 350–500 °C are mainly characterized by a relatively intact spreading and rebound process (see Fig. 3), and the experimental curves remain generally smooth, as shown in Fig. 6. For the 350–500 °C cases, the evolution trends of $\beta(t)$ at different temperatures are generally similar and all exhibit a behavior typical of rapid spreading after impact, followed by a retraction stage after reaching the maximum spreading diameter. In contrast, the curve at 300 °C shows an evolution trend distinct from those of the other temperature groups from the early stage onward, and does not exhibit the typical Leidenfrost spreading trajectory consistent with the prediction model in Ref. [26].

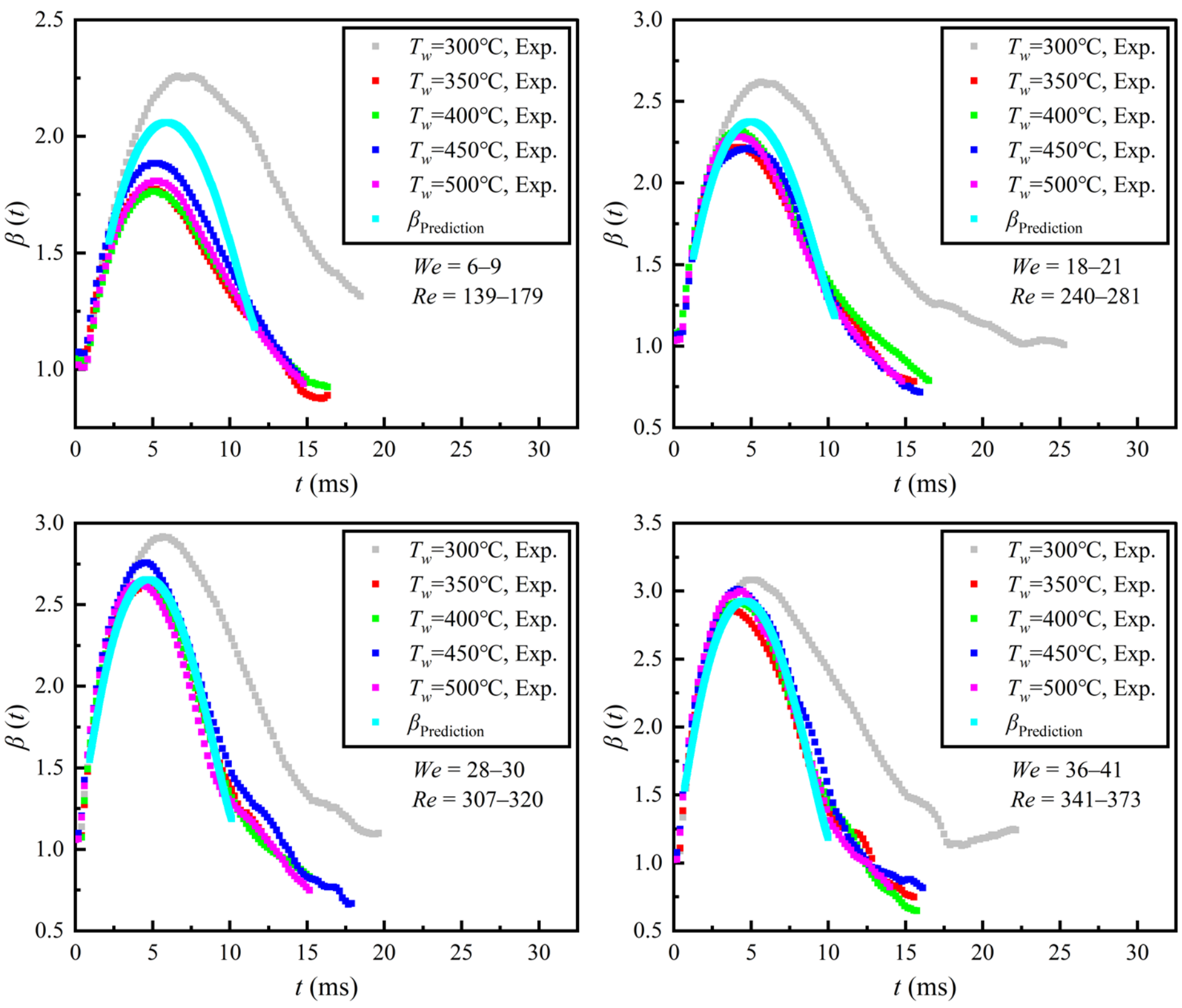

**Fig. 6.** Evolution of the droplet spreading factor, $\beta(t)$, under low-$We$ conditions. No droplet

breakup is observed under these conditions. The temperatures indicated in the legends are wall temperatures. The $\beta(t)$ < 1 values after rebound result from continued image tracking of vertically elongated droplets and should not be interpreted as near-wall spreading or retracting.

As shown in Fig. 6, when $We$ increases from approximately 8 to approximately 39, the experimental curves at wall temperatures of 350–500 °C can, overall, be well captured by the model in Ref. [26] both before and after the droplet reaches its maximum spreading, especially when $We > 20$. In terms of the curve shape depicted in Fig. 6, the model accurately predicts both the rapid spreading process driven by the initial inertia and the variations of $\beta(t)$ with $We$ and $Re$ during the retraction stage. This result indicates that, within the low-$We$ regime, droplet dynamics are still mainly governed by the competition among inertia, surface tension, and viscous dissipation, and the basic assumptions of the model in Ref. [26] regarding continuous liquid-film spreading can provide reasonable predictions of the experimental results within this range. Note that, at very low $We$, the model in Ref. [26] cannot predict $\beta(t)$ well, which is consistent with the observations reported in Ref. [26].

From the comparison among curves at different temperatures, the overall effect of temperature on $\beta(t)$ remains relatively limited in this regime. No obvious separation among the curves is observed during the spreading stage or near the maximum spreading point. This indicates that wall temperature has only a weak effect on $\beta(t)$ in the low-$We$ regime. In other words, under low-$We$ and Leidenfrost conditions, although wall temperature may influence the vapor-film state and some details of the later local recoil process, it is not yet sufficient to significantly alter the overall spreading dynamics of the droplet. This provides a useful reference for the stronger temperature dependence observed later in the intermediate-$We$ regime.

### 3.2 Transitional behavior in the intermediate-$We$ regime

Fig. 7 shows the temporal evolution of the spreading factor $\beta(t)$ in the intermediate-$We$ regime, where breakup begins to occur for the Leidenfrost cases at 350–500 °C, and the experimental results are compared with the model in Ref. [26]. Compared with the low-$We$ regime, the $\beta(t)$ curves from experiments at this stage, as shown in Fig. 7, still retain the overall spreading–retraction profile, but the differences among the curves at different wall temperatures are no longer limited to very slight deviations. Instead, they gradually exhibit clearer divergence and more pronounced shifts. This indicates that, with increasing impact inertia, droplet dynamics begin to enter a transitional regime that is more sensitive to temperature.

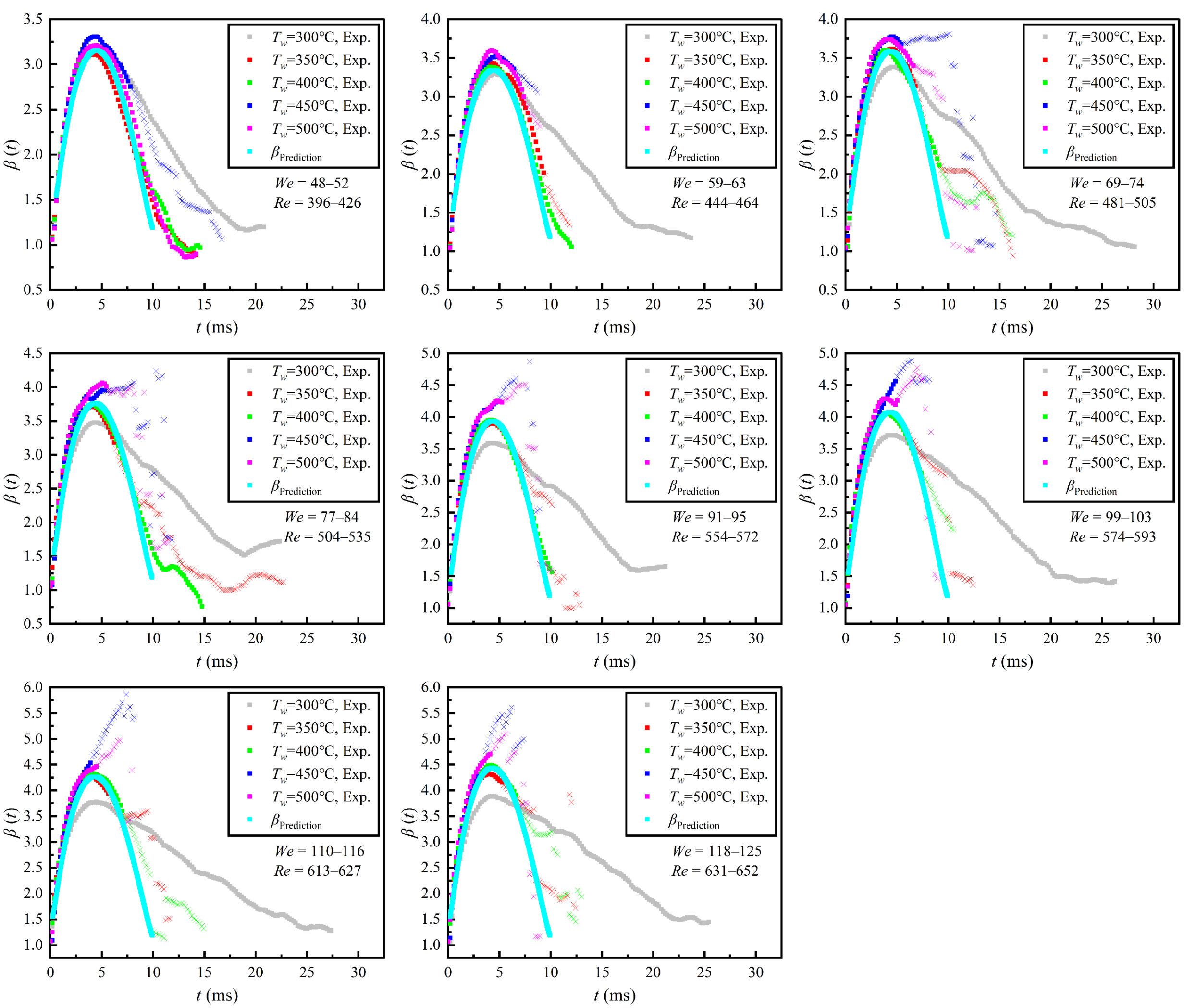

**Fig. 7.** Evolution of the droplet spreading factor, $\beta(t)$, under intermediate-$We$ conditions. For cases in which breakup is observed, the post-breakup experimental data are marked by "×" symbols.

As shown in Fig. 7, within the range of $We \approx$ 50–120, the experimental curves at 350–500 °C remain in good agreement with the model predictions of Ref. [26] during the initial spreading stage. In particular, within the first few milliseconds after impact, $\beta(t)$ and its variation near its peak under different temperature conditions are generally consistent with the predicted curves. This comparison indicates that, despite the temperature-dependent differences in breakup timing, the pre-breakup $\beta(t)$ evolution at different wall temperatures still follows the prediction model in Ref. [26]. Therefore, this model can still serve as a reasonable framework for describing the early-stage spreading dynamics.

However, unlike in Fig. 6, the experimental curves in Fig. 7 begin to show obvious divergence around the maximum spreading point. As shown in Fig. 7, the time at which the curves deviate from the model prediction is not the same at different temperatures, and the subsequent evolution trajectories also differ significantly. Under wall-temperature conditions of 450 and 500 °C, the experimental curves deviate rapidly at an earlier time, whereas under wall temperature conditions of 350 and 400 °C, deviation does not begin until a later stage. Fig. 7 shows that wall temperature does not significantly affect the droplet behavior before the maximum spreading is reached, whereas its effect on the subsequent evolution, especially on the onset time of breakup, is very pronounced. This result suggests that the temperature does not affect the early inertial spreading rate as strongly as it influences the evolution path of the experimental curves by affecting the late-stage thin-film state, the growth of interfacial disturbances, and the conditions that trigger breakup. In other words, within this $We$ regime, the droplet can still undergo rapid spreading dominated by inertia, but the liquid film has already become thin enough that the vapor-layer effect induced by the heated wall and local instabilities can be amplified in the later stage, ultimately leading to significant differences in breakup timing at different temperatures.

This result is also important for evaluating the applicability of the model. Fig. 7 does not imply that the model in Ref. [26] fails entirely in the intermediate-$We$ regime. On the contrary, it clearly shows that the model still has strong descriptive capability for the spreading process before breakup occurs. The real discrepancy mainly appears after the liquid film enters the unstable and breakup stages. Once the droplet enters the breakup-dominated stage, the assumption of a continuous liquid film is no longer valid, and the model can no longer accurately describe the subsequent experimental behavior. Therefore, if the research objective is the early-stage spreading dynamics of droplets impacting a heated wall, the model in Ref. [26] can still provide a reliable reference. However, if one further aims to predict the breakup time and the subsequent fragmentation behavior at different temperatures, the model must be modified to include additional instability or breakup criteria.

### 3.3 Splashing-dominated behavior in the high-$We$ regime

For the Leidenfrost cases at 350–500 °C, as $We$ continues to increase, the dynamical characteristics of droplet impact undergo another pronounced change. Under these conditions, the droplet possesses higher impact inertia, and the liquid film expands rapidly within a very short time and becomes thinner, while the later-stage instability also becomes more intense. Fig. 8 presents the evolutions of the droplet spreading factor at high $We$, where the droplet behavior gradually shifts from continuous liquid-film spreading to a breakup process dominated by splashing and fragmentation.

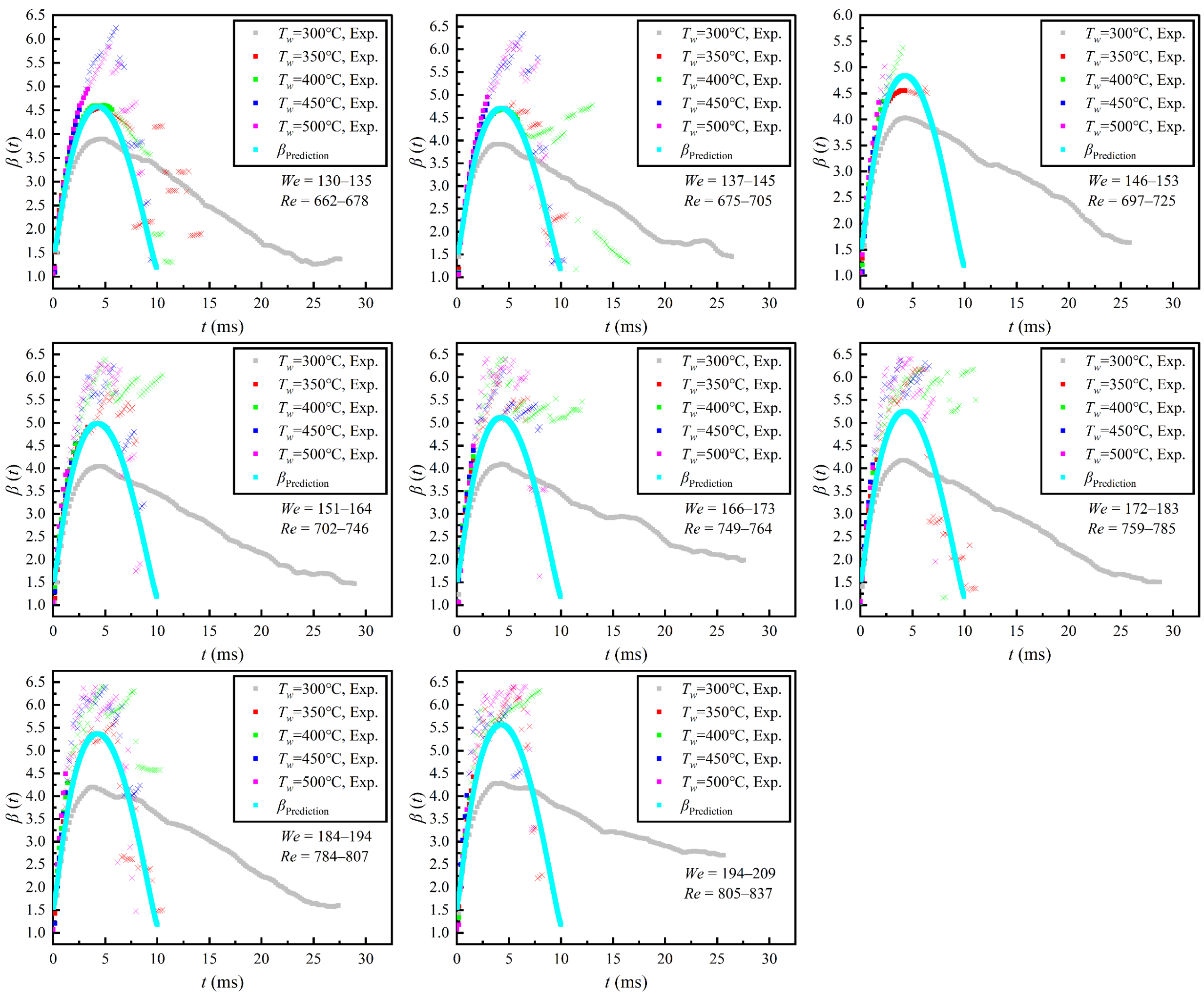

**Fig. 8.** Evolution of the droplet spreading factor, $\beta(t)$, under high-$We$ conditions. For cases in which breakup is observed, the post-breakup experimental $\beta(t)$ data are marked by "×" symbols.

Despite this shift in the dominant droplet behavior, the model in Ref. [26] still retains some descriptive capability before breakup occurs. It is worth noting that, before the droplet breaks up, the evolution of $\beta(t)$ still remains in close agreement with the model prediction. In the high-$We$ regime, the model can still, to some extent, capture the overall trend of early-stage spreading growth. In particular, before the droplet reaches the maximum spreading point, the predicted curves still show a certain degree of agreement with the experimental results. This indicates that, even under high-$We$ conditions, the early spreading stage can still be regarded as a short continuous-spreading process before the onset of strong breakup.

However, compared with the intermediate-$We$ regime, the deviation between the experimental curves and the model predictions in Ref. [26] appears earlier and becomes more pronounced in this range. As shown in Fig. 8, under high-$We$ conditions, the experimental curves deviate from the model prediction soon after reaching the maximum spreading point, and a large number of discrete scattered points distributed upward can be observed. Compared with Fig. 7, the most notable feature in Fig. 8 is not a further enhancement of the temporal separation among the curves at different temperatures, but rather that nearly all cases rapidly enter a state of obvious deviation from the prediction within a short period of time. In other words, when $We$ is sufficiently high, the later-stage droplet dynamics are no longer mainly characterized by subtle differences in when deviation begins, but by the common feature that almost all cases quickly enter a splashing-induced breakup state.

This rapid transition also changes the relative importance of wall temperature. Although some variations in the experimental data at different wall temperatures still exist, the breakup onset times corresponding to different temperatures in Fig. 8 do not show the strong separation that appears in Fig. 7. In many cases, obvious deviation from the continuous liquid-film model begins within a similar timescale. This indicates that, in the high-$We$ regime, the modulating effect of wall temperature on breakup timing becomes relatively weaker, whereas impact inertia becomes the primary factor governing the later-stage droplet behavior. From a physical point of view, this phenomenon indicates that the dominant mechanism gradually shifts from joint control by inertia and thermal effects to inertia-dominated behavior. At high $We$, the rim edge and the thin-liquid-film edge that forms after impact have higher velocity and stronger instability, so that local disturbances can grow rapidly within a very short time and develop into splashing and fragmentation. Therefore, the relative convergence of breakup timing at different temperatures in Fig. 8 essentially reflects the strengthening of inertial control at high $We$.

Once breakup occurs, the measured $\beta(t)$ no longer represents a smooth and continuous spreading or retraction process. The discrete scattered points observed after breakup no longer correspond to the retraction of a continuous lamella–rim structure, but instead reflect significant breakup and secondary droplet ejection in the later stage. Therefore, once the droplet enters the clearly splashing- and fragmentation-dominated stage, the assumption of a continuous liquid film is no longer valid, and the model in Ref. [26] can no longer accurately describe the subsequent experimental behavior. This means that, for high-$We$ conditions, the model in Ref. [26] is more appropriate as a reference model for the early spreading stage than as a predictor of the detailed dynamics of the later breakup-dominated stage.

### 3.4 Comparison of results across different $We$ regimes

For the Leidenfrost cases at 350–500 °C, Figs. 6–8 show that $We$ not only changes the spreading extent and timescale of the droplet, but also determines how the effect of wall temperature manifests itself throughout the entire dynamic process. In the low-$We$ regime, the droplet generally remains intact, with behavior still dominated by spreading and retraction; the experimental curves remain relatively smooth, the differences among the curves at different temperatures are limited, and the model in Ref. [26] shows good applicability to the overall spreading process. As $We$ increases into the intermediate range, thinning of the liquid film accelerates, and late-stage instability begins to intensify. Under these conditions, the experimental curves at different temperatures exhibit clear divergence around the breakup stage, and the influence of wall temperature on breakup timing becomes particularly pronounced. With a further increase in $We$, the droplet rapidly shifts to a splashing-dominated state, the experimental curves rapidly deviate from the model prediction, and the breakup onset times at different temperatures become closer to one another, indicating inertia-dominated behavior.

Therefore, what these three groups of figures reveal is not simply whether the model fits well or poorly, but rather a dynamic picture that evolves progressively with $We$. Under low-$We$ conditions, the model in Ref. [26] can predict the main spreading behavior of the droplet reasonably well. Under intermediate-$We$ conditions, the model remains valid for the pre-breakup dynamics, while the modulation of breakup timing by temperature becomes very significant. Under high-$We$ conditions, the later-stage droplet behavior rapidly shifts to an inertia-dominated splashing breakup process, with the deviation between the model prediction and the experiments mainly concentrated after the maximum spreading point. These results indicate that the model in Ref. [26] is suitable as a fundamental descriptive framework for the pre-breakup spreading dynamics of droplets impacting heated walls, whereas the variation of breakup onset with $We$ under different temperature conditions requires further discussion in conjunction with liquid-film instability and breakup mechanisms.

**3.5 Overall characteristics of the temperature effect**

Figs. 6–8 show that the effect of wall temperature does not simply scale with either the spreading factor or the breakup time. Instead, its role depends on the stage of impact and on whether the wall temperature is above the Leidenfrost point. The 300 °C cases are under non-Leidenfrost conditions. Their behavior shows that crossing the Leidenfrost threshold

fundamentally changes the impact mode. In this sense, temperature first acts as a state-discriminating parameter: it determines whether the droplet can enter a vapor-layer-supported Leidenfrost impact state, in which the continuous lamella-spreading model becomes comparable with the experiments.

Once the Leidenfrost state is established, the influence of wall temperature on the early spreading stage is limited. For the 350–500 °C cases, the $\beta(t)$ curves at different wall temperatures nearly overlap before the maximum spreading point. This indicates that the early spreading rate and spreading trajectory are mainly determined by impact inertia. Wall temperature does not significantly modify this inertial spreading process.

The temperature effect becomes more important in the later stage, when the lamella becomes thin and its stability becomes more sensitive to local conditions. In this stage, wall temperature can influence the vapor-layer state, the momentum exchange between the liquid film and the heated wall, and the amplification of local disturbances. Therefore, once Leidenfrost-regime behavior is established, temperature acts mainly as an influencing factor for instability onset and subsequent breakup, rather than as a parameter that reshapes the early spreading motion. Overall, the temperature effect is hierarchical: for cases that do not enter the Leidenfrost regime, it determines the basic impact state; for cases that exhibit Leidenfrost-regime behavior, it mainly affects the late-stage lamella stability and the sensitivity of breakup onset.

### 3.6 Assessment of local indicators for breakup onset

Before using lamella thickness to interpret breakup timing, the dimensionless rim-center velocity, $\dot{R}_R^*$, is first examined as a possible breakup indicator. The dimensionless velocity, $dR_R^*/dt^*$ ($\dot{R}_R^*$), is obtained as part of the numerical integration of Eq. (21), which is the governing equation for $R_R^*(t^*)$ in the model of Ref. [26]. In the numerical implementation, Eq. (21) is solved as a first-order system with $R_R^*$ and $\dot{R}_R^*$ as the two state variables. Fig. 9 shows the theoretical evolution of $dR_R^*/dt^*$ under different Weber number conditions. The experimentally identified breakup instants are also mapped onto the corresponding velocity curves.

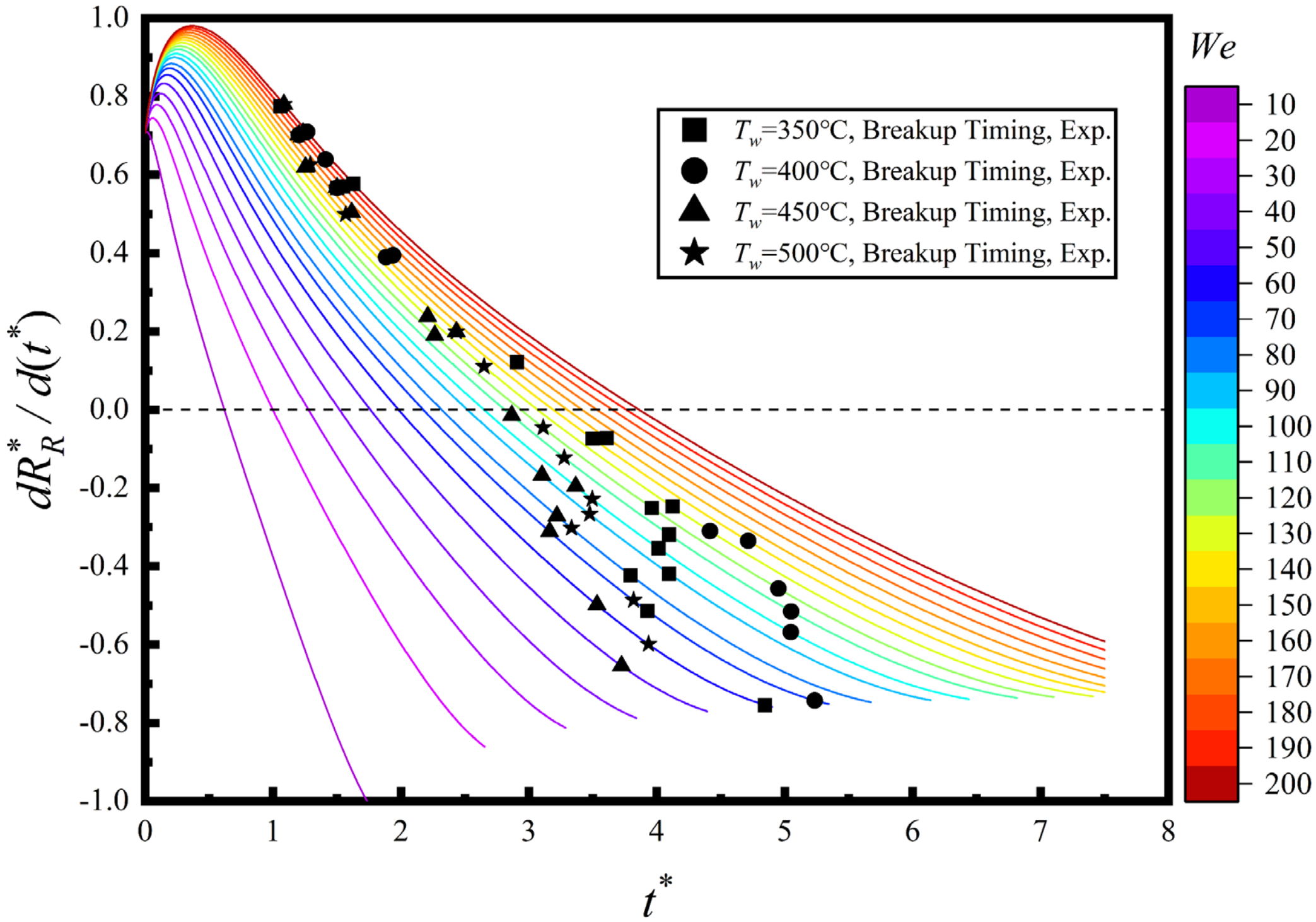

**Fig. 9.** Dimensionless rim velocity evolution under different $We$ conditions and corresponding breakup timing. The colored curves denote the model-predicted evolution of $dR_R^*/dt^*$, and the symbols indicate the experimentally identified breakup instants at different wall temperatures.

The velocity curves show the rim-center motion from outward spreading to inward retraction. If the rim-center radial velocity, $\dot{R}_R$, controlled breakup onset, the breakup points would be expected to collapse into a narrow velocity range. However, the breakup points are broadly dispersed in the $t^* - dR_R^*/dt^*$ plane. They do not form a clear velocity band. This distribution is different from the lamella-thickness plot discussed later, where the breakup points are mainly confined within a narrow low-$h_R^*$ range. Therefore, $dR_R^*/dt^*$ alone is not a robust indicator of breakup onset. The velocity describes the motion state of the lamella-rim region, but it does not provide a clear critical value for breakup. Since this velocity does not give a clear breakup criterion, the rim-center radial acceleration, $\ddot{R}_R^*$, is examined next.

The acceleration associated with $R_R$ is also examined to determine whether it can provide a clearer indication of breakup onset. Fig. 10 shows the theoretical evolution of the dimensionless acceleration, $d^2R_R^*/d(t^*)^2$ ($\ddot{R}_R^*$), under different Weber number conditions. This acceleration is numerically calculated from Eq. (21). The experimentally identified breakup instants are mapped onto the corresponding acceleration curves. The breakup points approximately follow a linear trend in the $t^* - d^2R_R^*/d(t^*)^2$ plane. However, this trend does not indicate a clear acceleration-based breakup criterion.

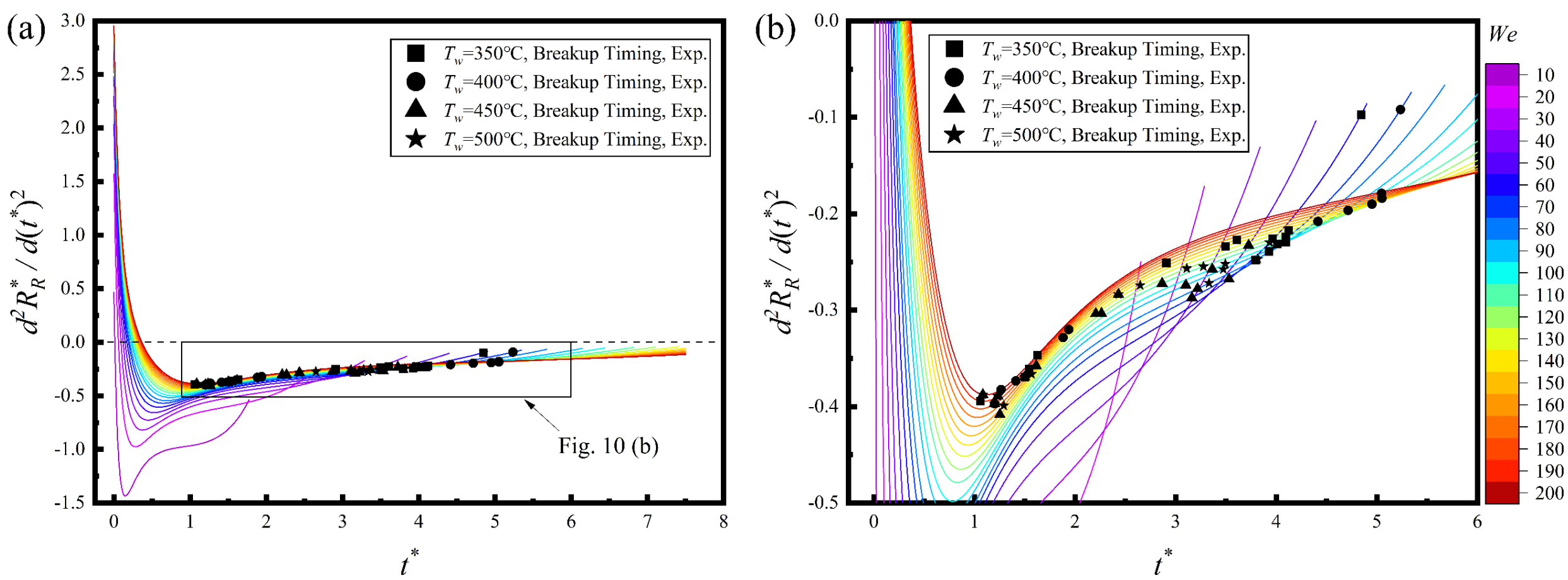

**Fig. 10.** Dimensionless rim acceleration evolution under different $We$ conditions and corresponding breakup timing. The colored curves denote the model-predicted evolution of $d^2R_R^*/(dt^*)^2$, and the symbols indicate the experimentally identified breakup instants at different wall temperatures. Panel (a) shows the full acceleration evolution, while panel (b) provides an enlarged view of the boxed region in panel (a), where most breakup points are located. The dashed horizontal line indicates zero acceleration.

More specifically, all breakup points are located in the negative-acceleration region. This means that breakup mainly occurs after the motion of $R_R$ has entered a deceleration stage. However, the breakup points do not appear in the high positive-acceleration region, where the lamella-rim region is rapidly accelerated outward. They also do not cluster near the negative acceleration extrema, namely the maximum-deceleration points. Therefore, breakup onset is not associated with either a strong positive acceleration or a characteristic maximum deceleration.

These results indicate that $d^2R_R^*/(dt^*)^2$ is not a robust standalone indicator of breakup onset. The acceleration plot can identify the general dynamic stage in which breakup occurs. The breakup points are not located in the high positive-acceleration region, and they also do not cluster near the extrema of the curves. Therefore, the analysis next turns from the kinematic quantities of $R_R$ to the local lamella state.

Fig. 11 presents the model-predicted evolution of the dimensionless lamella thickness at the lamella–rim junction, $h_R^* = h_{\mathrm{lam}}^*(R_R^*, t^*)$, under different Weber number conditions. The results are plotted as a function of dimensionless time, $t^*$, together with the experimentally identified breakup instants mapped onto the $t^* - h_R^*$ plane. The colored curves denote the evolution of $h_R^*$ at different Weber numbers, while the symbols represent the corresponding values of $h_R^*$ and $t^*$ at breakup under different wall temperature conditions. The solid line indicates a linear fit to the breakup points, and the shaded regions represent the corresponding 95% prediction intervals.

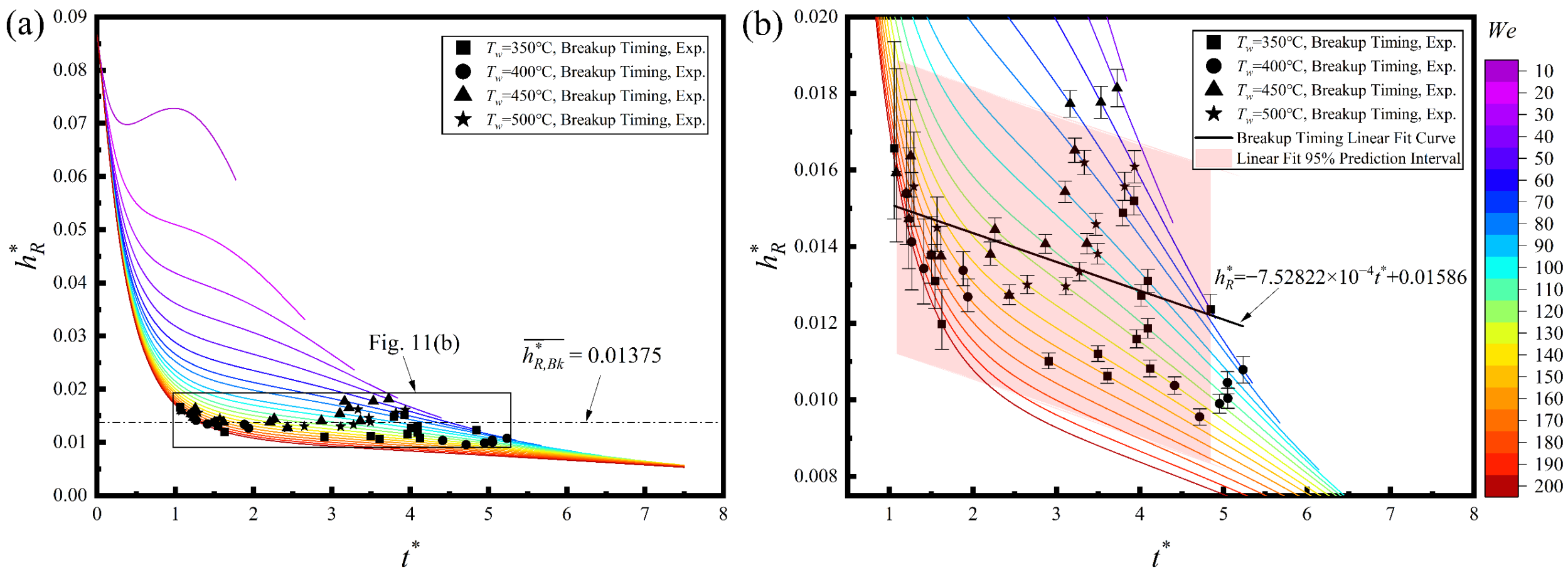

**Fig. 11.** Dimensionless lamella thickness evolution under different *We* conditions and corresponding breakup timing. The colored curves denote the model-predicted $h_R^*$, and the symbols indicate the experimentally identified breakup instants at different wall temperatures. Panel (a) shows the overall evolution, where $\overline{h_{R,Bk}^*}$ denotes the average $h_R^*$ value of the breakup instants; panel (b) enlarges the breakup region with the linear fit and the corresponding 95% prediction intervals.

Based on this mapping, a key feature in Fig. 11(a) is that breakup does not occur at a fixed dimensionless time. Instead, the breakup points are mainly associated with the thinning state of the lamella. This suggests that breakup onset is more closely related to whether the lamella has entered a sufficiently thin, instability-prone range than to a common value of $t^*$. At the experimentally identified breakup instants, the model-inferred dimensionless lamella thickness has an average value of $\overline{h_{R,Bk}^*} = 0.01375$. In other words, the global dimensionless time alone is not sufficient to determine the onset of breakup.

More specifically, as shown in Fig. 11(b), the breakup points are mainly confined within a relatively narrow lamella-thickness range, approximately $h_R^* \approx$ 0.010–0.017. This range is determined as the approximate 5th–95th percentile interval of the model-inferred $h_{R,Bk}^*$ values from all experimentally identified breakup cases. This indicates that breakup tends to occur after the lamella thins into this range, where local disturbances can be amplified more readily. Therefore, the experimentally identified breakup points suggest that lamella thinning is closely related to breakup onset, rather than breakup being governed only by the global impact condition or by time alone.

The role of the Weber number can then be interpreted through its effect on the lamella-thinning rate. The Weber number primarily controls how rapidly the droplet approaches this breakup-prone thickness window. As $We$ increases, the lamella thins more rapidly, and breakup correspondingly occurs at smaller $t^*$. Therefore, higher-$We$ cases reach the breakup-prone thickness window earlier, which explains why the onset of breakup shifts to earlier dimensionless times as $We$ increases.

By contrast, for cases at the same $We$ value but different wall temperatures, breakup does not occur at the same $t^*$, leading to slight variations in the corresponding $h_R^*$ values. As a result, the breakup points form a finite band rather than collapsing onto a single line. This indicates that breakup is better understood as occurring within an instability window, rather than at a single critical thickness. In this sense, wall temperature affects the exact breakup timing within this thickness range, but it does not change the overall tendency for breakup to occur after the lamella has become sufficiently thin.

Therefore, the significance of Fig. 11 lies not only in identifying the lamella-thickness range associated with breakup, but also in linking the low-, intermediate-, and high-$We$ results through a common physical picture. In the low-$We$ regime, the lamella generally remains too thick to enter the breakup-prone thickness window within the measurement timescale, so temperature has only a limited influence on breakup tendency. In the intermediate-$We$ regime, the lamella can thin into this range while thermal effects remain important, making the breakup timing more sensitive to wall temperature. In the high-$We$ regime, however, the lamella reaches the breakup-prone thickness window so rapidly that the subsequent evolution becomes increasingly inertia-dominated. These results suggest that $h_R^*(t^*)$ can serve as a useful bridge between continuous spreading behavior and breakup-onset criteria.

The present analysis also suggests a possible way to develop a data-informed breakup criterion for wall-impinging droplets. For a given liquid, such as *n*-hexadecane, a sufficiently large experimental database could be used to obtain the probability distribution of breakup onset within the breakup-prone $h_R^*$ window under different surface-temperature conditions. In practical spray-wall simulations, Eq. (21) can be numerically integrated to obtain $R_R^*(t^*)$, and Eq. (17) can then be used to calculate the corresponding $h_R^*(t^*)$ trajectory. When this trajectory enters the breakup-prone thickness window, the experimentally obtained probability distribution could be used to estimate the likely breakup onset time. Compared with conventional empirical droplet breakup models based mainly on global parameters such as $We$ and $Oh$ (Ohnesorge number, $Oh = \sqrt{We}/Re$), which predict whether a single droplet breaks up [34], the approach of this study provides additional temporal and local-state information for describing droplet breakup.

## 4. Conclusions

The impact and breakup behaviors of single *n*-hexadecane droplets on a heated stainless-steel wall are experimentally investigated under different wall-temperature and Weber number conditions, and the applicability of the lamella-rim model reported in Ref. [26] to the pre-breakup spreading process is assessed. The results show that droplet impact behavior is controlled by both impact inertia and wall temperature, but that their roles are not equivalent throughout the evolution process.

Within the present experimental range, the overall impact outcome changes progressively with increasing Weber numbers. At 300 °C, all tested cases exhibit the deposition behavior, indicating that this condition did not exhibit stable Leidenfrost impact behavior under the present impact conditions. By contrast, for the 350–500 °C cases, the impact behavior evolves from rebound to ejection, fragmentation, and splashing as the Weber number increases. This indicates that Weber number primarily determines the overall transition sequence of impact behavior, whereas wall temperature shifts the Weber-number ranges associated with the Leidenfrost outcomes.

Comparison with the lamella-rim model reported in Ref. [26] further shows that the model retains clear value as a description of the continuous spreading stage prior to breakup. In the low-$We$ regime for the Leidenfrost cases at 350–500 °C, the predicted spreading-factor evolution agrees well with the experimental results over most of the spreading and retraction process. In the intermediate-$We$ regime, the model accurately represents the early spreading dynamics and the evolution near maximum spreading, but deviates from the experimental results as breakup begins. In the high-$We$ regime, the model remains useful mainly as a reference for the early spreading stage, since the later evolution rapidly becomes dominated by splashing and fragmentation. These results indicate that the model in Ref. [26] is suitable for describing pre-breakup continuous spreading, but not for predicting the subsequent breakup process itself.

The wall-temperature effect is strongly stage-dependent. The 300 °C non-Leidenfrost cases demonstrate that crossing the Leidenfrost threshold fundamentally changes the droplet impact mechanism. For the Leidenfrost cases at 350–500 °C, however, wall temperature has only a limited influence on the early spreading process, and the $\beta(t)$ curves before the maximum spreading point remain largely similar under different temperature conditions. The wall-temperature effect becomes much more evident in the later stage, where it modulates breakup timing after lamella thinning. This temperature dependence is most pronounced in the intermediate-$We$ regime, whereas in the high-$We$ regime the relative importance of wall temperature decreases as the breakup process becomes increasingly inertia-dominated.

Analysis of the lamella-thickness evolution provides a more unified interpretation of the breakup results. The model-inferred lamella thicknesses evaluated at experimentally observed breakup instants are mainly concentrated between 0.010 and 0.017 of the initial droplet diameter, rather than being associated with a fixed dimensionless time. Under the present conditions ($n$-hexadecane droplets, a stainless-steel wall, atmospheric pressure, and the present $We$ range), this result suggests that lamella thinning provides a more useful local-state indicator of breakup timing than dimensionless time alone. Weber number primarily controls how rapidly the lamella approaches this breakup-prone thickness window, while wall temperature modulates the breakup timing after the lamella has thinned.

The proposed thickness window should not be interpreted as a universal critical value. It is obtained from a model-assisted reconstruction of the lamella state for $n$-hexadecane droplets under the present wall material, wall temperature conditions, ambient pressure, and Weber-number conditions. Direct lamella-thickness measurements and a broader range of liquids and wall/surface conditions are needed before the criterion can be generalized. Nevertheless, the present results suggest that model-inferred lamella thickness can complement conventional Weber-number and wall-temperature regime maps by providing local-state information for breakup timing in hot-wall fuel-droplet impingement models.

**CRediT authorship contribution statement**

Xinquan Liang: Conceptualization, Methodology, Investigation, Data curation, Formal analysis, Software, Visualization, Writing – original draft.

Daniel Olufemi Olurotimi: Writing – review & editing.

Xi Liu: Writing – review & editing.

Minghui Xu: Writing – review & editing.

Yuhao Xu: Conceptualization, Methodology, Validation, Supervision, Project administration, Resources, Funding acquisition, Writing – review & editing.

**Data availability**

The processed data supporting this study will be made available by the corresponding author upon reasonable request. Additional raw image/video data will also be made available from the corresponding author upon reasonable request.

**Declaration of generative AI and AI-assisted technologies in the writing process**

During the preparation of this work, the authors used ChatGPT to check and improve spelling and grammar. After using this tool/service, the authors reviewed and edited the content as needed and take full responsibility for the content of the publication.

**Declaration of competing interest**

The authors declare that they have no known competing financial interests or personal relationships that could have appeared to influence the work reported in this paper.